\documentclass[preprint,amsmath,amssymb,11pt,english,aps,showkeys,showpacs]{revtex4}
\usepackage[english]{babel}
\usepackage{graphicx}
\usepackage{graphics}
\usepackage{amsmath}
\usepackage{dcolumn}
\usepackage{amssymb}
\usepackage{bm}

\begin{document}
\title{Geometric quantum phases from Lorentz symmetry breaking effects in the cosmic string spacetime}
\author{H. Belich} 
\email{belichjr@gmail.com}
\affiliation{Departamento de F\'isica e Qu\'imica, Universidade Federal do Esp\'irito Santo, Av. Fernando Ferrari, 514, Goiabeiras, 29060-900, Vit\'oria, ES, Brazil.}

\author{K. Bakke}
\email{kbakke@fisica.ufpb.br}
\affiliation{Departamento de F\'isica, Universidade Federal da Para\'iba, Caixa Postal 5008, 58051-970, Jo\~ao Pessoa, PB, Brazil.}

\begin{abstract}
By starting from the modified Maxwell theory coupled to gravity, the arising of geometric quantum phases in the relativistic and nonrelativistic quantum dynamics of a Dirac neutral particle from the effects of the violation of the Lorentz symmetry in the cosmic string spacetime is investigated. It is shown that the Dirac equation can be written in terms of an effective metric and a relativistic geometric phase stems from the topology of the cosmic string spacetime and the Lorentz symmetry breaking effects. Further, the nonrelativistic limit of the Dirac equation is discussed and it is shown that both Lorentz symmetry breaking effects and the topology of the defect yields a phase shift in the wave function of the nonrelativistic spin-$1/2$ particle. 
\end{abstract}


\maketitle

\section{Introduction}

With the recent discovery of the Higgs boson at the LHC a research program that aims to explain the physics of fundamental interactions as the manifestation of excitations of fundamental fields is completed. With this discovery, it closes a scenario of forecasts of experimental success. Despite this tremendous progress, it is known that the description neutrino without mass is unsatisfactory. Also, there is a lack of theory which explains the microscopic origin of the boson that generates mass to all particles. However, the Standard Model (SM) and General Relativity are the basis of our understanding of the physics of the fundamental interactions. In fact, both are effective theories and we hope that new physics may appear when the LHC reaches the 14 TeV, or goes beyond. As effective theories, we might question what concepts that could be used as guides for a fundamental theory. Taking into account the concepts which lead to the SM electroweak unification, the Higgs mechanism is an interesting way to get clues about how to obtain a fundamental theory guide. Thereby, we can describe a phase transition without elaborating on the microscopic level and, most importantly, the gauge symmetry is preserved.

At present days, it is well-known that the Standard Model (SM) of particle physics has not gotten success of explaining the origin of electron electric dipole moment (EDM), $d_{e}$, and its experimental upper bounds \cite{revmod}. Theories beyond the Standard Model predict a small, but potentially measurable EDM ($d_{e}\leq10^{-29}\,\mathrm{e}\cdot\mathrm{cm}$) \cite{science}, which presents an asymmetric charge distribution along the spin axis. Therefore, with this experimental result, it is necessary to investigate the physics beyond the Standard Model. Kosteleck\'y and Stuart \cite{extra3,extra3a,pot} showed that the interactions in the context of the string field theory could lead to the spontaneous breaking of the Lorentz symmetry \cite{extra3, extra3a}. It has been shown in string field theory that it can also break the CPT-symmetry spontaneously \cite{pot}. This line of research has suggested that experiments on the kaon interferometry would be promising for the search for possible signs of the CPT-violation due to its high sensitivity.

In order to investigate interference effects in processes involving fermions, the introduction of a nonminimal coupling has been proposed with a direct relation to experimental interference measurements through the modified Dirac theory \cite{belich,belich1,belich3,bbs2,bb,petrov,petrov2}. An interesting case is the spectrum of energy of the hydrogen atom discussed in Refs. \cite{Nonmini,novo,novo1} by considering the nonrelativistic limit of the modified Dirac theory. Besides, a great deal of works in the literature has looked at the Lorentz symmetry violation and numerous experimental bounds have been obtained \cite{extra2}. Experiments involving geometric quantum phases \cite{ab,ahan,anan,anan2,berry,anan3,tg1} are also interesting scenarios which introduce non-minimal couplings with a background of the Lorentz symmetry violation \cite{bbs3}. Other empirical studies of the CPT/Lorentz-violation have been made by considering the Minkowski spacetime such as involving muons \cite{muon}, mesons \cite{meson,meson2}, baryons \cite{barion}, photons \cite{photon}, electrons \cite{electron}, neutrinos \cite{neutrino} and the Higgs sector \cite{higgs}. The gravity sector has also been explored in \cite{gravity,gravity2,curv}. In Ref. \cite{data}, one can find the current limits of the coefficients for the Lorentz violation symmetry.

 Here, we deal with a Lorentz symmetry violating tensor $K_{\mu\nu\kappa\lambda}$ in such a way that the CPT symmetry is preserved. In recent years, it has been shown in Ref. \cite{hugo} that a particular decomposition of the tensor $K_{\mu\nu\kappa\lambda}$ produces a modification on the equations of motion of the electromagnetic waves due to the presence of vacuum anisotropies, which gives rise to the modified Maxwell equations. As a consequence, the anisotropy can be a source of the electric field and, then, Gauss's law is modified. Besides, the Amp\`ere-Maxwell law is modified by the presence of anisotropies and it has a particular interest in the analysis of vortices solutions since it generates the dependence of the vortex core size on the intensity of the anisotropy. 
 
It is well-known that the presence of terms that violate the Lorentz symmetry imposes at least one privileged direction in the spacetime. Another interesting effect that appears from the propagation of electromagnetic waves in a material medium is the Faraday effect (turning the polarization plane of the electromagnetic wave); this effect emerges as a property of the vacuum when it imposes a term in which, apart from violating the Lorentz symmetry and modifying the speed of electromagnetic wave propagation, the CPT symmetry is also violated \cite{baeta}. On the other hand, it has been shown in Ref. \cite{hugo} that the violation of the Lorentz symmetry given by the tensor background defined by $K_{\mu\nu\kappa\lambda}$ modifies the Maxwell equations with sources in such a way that the dispersion relations are modified without the Faraday effect. In this case, the speed of the light in the vacuum can be changed without turning the plane of polarization of the light.

In this work, we analyse the arising of geometric quantum phases in the relativistic and nonrelativistic quantum dynamics of a Dirac neutral particle from the effects of the violation of the Lorentz symmetry in the cosmic string spacetime. By starting from the modified Maxwell theory coupled to gravity discussed in Refs. \cite{curv2,curv3}, we establish a spacelike normalized parameter four-vector $\xi^{a}$ and write an effective metric for the cosmic string spacetime under Lorentz symmetry breaking effects. Then, we write the Dirac equation based on this effective metric and obtain the relativistic geometric phase which stems from the topology of the cosmic string spacetime and the Lorentz symmetry breaking effects. We also discuss the nonrelativistic limit of the Dirac equation and show that both Lorentz symmetry breaking effects and the topology of the defect yields a phase shift in the wave function of the nonrelativistic spin-$1/2$ particle.

This paper is organized as follows: in section II, we start by making a brief introduction the spinor theory in curved space and the modified Maxwell theory coupled to gravity. In the following, we establish a spacelike normalized parameter four-vector $\xi^{a}$ and write an effective metric for the cosmic string spacetime under Lorentz symmetry breaking effects. Then, we discuss the arising of geometric quantum phases in the wave function of a Dirac particle that stems from the topology of the cosmic string and the Lorentz symmetry breaking effects; in section III, we discuss the nonrelativistic limit of the Dirac equation by applying the Foldy-Wouthuysen approximation \cite{fw,greiner} and obtain the geometric quantum phase yielded by Lorentz symmetry breaking effects and the topology of the defect; in section IV, we present our conclusions.

\section{Fermions in presence of an effective metric and relativistic geometric quantum phases}

In this section, we start from the modified Maxwell theory coupled to gravity in order to establish a spacelike normalized parameter four-vector $\xi^{a}$ and write an effective metric for the cosmic string spacetime under Lorentz symmetry breaking effects. Then, we show that geometric quantum phases can arise from the effects of the Lorentz symmetry breaking effects and the topology of the cosmic string spacetime.

The extension of the Standard Model (SME) has been a source of intense investigations by not presenting the usual changes in the QED, such as the violation of the CPT-symmetry and birefringence in vacuum. The CPT-even gauge sector of the SME has been studied since 2002, after the pioneering contributions made by Kosteleck\'y and Mewes \cite{19,20}, and, at present days, there is an extensive literature dealing with the extension of the Standard Model in the even sector of SME by the following term \cite{cas}:
\begin{equation}
S=-\frac{1}{4}\int d^{4}x\;K_{abcd}\,F^{ab}\,F^{cd}. 
\label{1.1}
\end{equation}

The tensor $K_{abcd}$ is CPT-even, that is, it does not violate the CPT-symmetry. Although the violation of the CPT-symmetry implies that the Lorentz invariance is violated \cite{greenberg}, the reverse is not necessarily true. The action given in Eq. (\ref{1.1}) breaks the Lorentz symmetry in the sense that the tensor $K_{abcd}$ has a non-null vacuum expectation value. Besides, the tensor $K_{abcd}$ has the same properties of the Riemann tensor, as well as an additional double-traceless condition. This tensor possesses the following symmetries:
\begin{equation}
K_{abcd}=K_{\left[ab\right]\left[cd\right]};\,\,\,K_{abcd}=K_{cdab};\,\,\,K^{ab}_{\,\,\,\,\,\,\,ab}=0.
\label{1.2}
\end{equation}

By following Refs. \cite{curv3,curv2,curv4}, we can write the tensor $K_{abcd}$ in terms of a traceless and symmetric matrix $\tilde{\kappa}_{ab}$ as
\begin{equation}
K_{abcd}=\frac{1}{2}\left(\eta_{ac}\,\tilde{\kappa}_{bd}-\eta _{ad}\,\tilde{\kappa}_{bc}+\eta_{bd}\,\tilde{\kappa}_{ac}-\eta_{bc}\tilde{\kappa}_{ad}\right).
\label{1.3}
\end{equation}

By defining a normalized parameter four-vector $\xi^{a}$, which satisfies the conditions: $\xi_{a}\xi^{a}=1$ for the timelike case and $\xi_{a}\xi^{a}=-1$ for the spacelike case; thus, we can decompose the tensor $\tilde{\kappa}_{ab}$ as 
\begin{equation}
\tilde{\kappa}_{ab}=\kappa\left(\xi_{a}\xi_{b}-\frac{\eta_{ab}\,\xi ^{c}\xi_{c}}{4}\right),
\label{1.4}
\end{equation}
where $\kappa=\frac{4}{3}\tilde{\kappa}^{ab}\,\xi_{a}\,\xi_{b}$. As in Refs. \cite{curv3,curv4}, we consider the parameter $\kappa$ being spacetime independent, where $0\leq\kappa<2$.

Recently, an interesting study has been conducted in a curved spacetime background where it is observed that the event horizon of a black hole is modified by the anisotropy generated by $K_{\mu\nu\kappa\lambda}$ \cite{curv2,curv3}. Therefore, an interesting way of analysing new phenomenology for verification Lorentz symmetry violation is to consider a curved spacetime background \cite{curv}. From this perspective, let us write the corresponding Lagrange density to the nonbirefringent modified Maxwell theory coupled to gravity \cite{curv2,curv3}:
\begin{equation}
\mathcal{L}_{\mathrm{modM}}=-\sqrt{g}\,\left(\frac{1}{4}\,F_{\mu\nu}F_{\rho\sigma}\,g^{\mu\rho}g^{\nu\sigma}+\frac{1}{4}\,K^{\mu\nu\rho\lambda}\,F_{\mu\nu}F_{\rho\lambda}\right).
\label{1.5}
\end{equation}

By using Eqs. (\ref{1.2}) and (\ref{1.3}), we can write the Lagrange density (\ref{1.5}) in terms of an effective metric tensor $\bar{g}_{\mu\nu}\left(x\right)$ as
\begin{eqnarray}
\mathcal{L}_{\mathrm{modM}}=-\sqrt{g}\left(1-\frac{1}{2}\,\kappa\,\xi_{\alpha}\,\xi^{\alpha}\right)\,\frac{1}{4}\,F^{\mu\nu}\left(x\right)F^{\rho\sigma}\left(x\right)\,\bar{g}_{\mu\rho}\left(x\right)\,\bar{g}_{\nu\sigma}\left(x\right),
\label{1.6}
\end{eqnarray}
where the expression of this effective metric tensor is given by \cite{curv2,curv3}:
\begin{eqnarray}
\bar{g}_ {\mu\rho}\left(x\right)=g_{\mu\rho}\left(x\right)-\epsilon\,\xi_{\mu}\,\xi_{\rho},
\label{1.7}
\end{eqnarray}
whose parameter $\epsilon$ is defined as $\epsilon=\frac{\kappa}{1+\frac{\kappa}{2}}$ and $\bar{g}^{\mu\nu}\,\bar{g}_{\nu\alpha}=\delta^{\mu}_{\,\,\,\alpha}$. However, all lowering or raising of indices is performed by using the original background metric $g_{\mu\nu}\left(x\right)$ and its inverse $g^{\mu\nu}\left(x\right)$. This type of background yields the anisotropy in the spacetime \cite{curv}, hence, this suggests that the propagation of the photon must be modified by this background.

Henceforth, let us discuss the behaviour of fermions in the cosmic string spacetime under Lorentz symmetry breaking effects. By considering a curved spacetime background, we deal with Dirac spinors by using the mathematical formulation of the spinor theory in curved space \cite{weinberg,bd}. Spinors are defined locally, where each spinor transforms according to the infinitesimal Lorentz transformations, that is, $\psi'\left(x\right)=D\left(\Lambda\left(x\right)\right)\,\psi\left(x\right)$, where $D\left(\Lambda\left(x\right)\right)$ corresponds to the spinor representation of the infinitesimal Lorentz group and $\Lambda\left(x\right)$ corresponds to the local Lorentz transformations \cite{weinberg,bd}. Locally, the reference frame of the observers can be build via a noncoordinate basis $\hat{\theta}^{a}=e^{a}_{\,\,\,\mu}\left(x\right)\,dx^{\mu}$, where the components $e^{a}_{\,\,\,\mu}\left(x\right)$ are called \textit{tetrads} and satisfy the relation: $g_{\mu\nu}\left(x\right)=e^{a}_{\,\,\,\mu}\left(x\right)\,e^{b}_{\,\,\,\nu}\left(x\right)\,\eta_{ab}$ \cite{weinberg,bd,naka}, where $\eta_{ab}=\mathrm{diag}(+ - - - )$ is the Minkowski tensor. The inverse of the tetrads are defined as $dx^{\mu}=e^{\mu}_{\,\,\,a}\left(x\right)\,\hat{\theta}^{a}$, and the relations $e^{a}_{\,\,\,\mu}\left(x\right)\,e^{\mu}_{\,\,\,b}\left(x\right)=\delta^{a}_{\,\,\,b}$ and $e^{\mu}_{\,\,\,a}\left(x\right)\,e^{a}_{\,\,\,\nu}\left(x\right)=\delta^{\mu}_{\,\,\,\nu}$ are satisfied.

The Dirac equation is defined in a curved spacetime background by considering the covariant derivative of a spinor, whose components are defined as $\nabla_{\mu}=\partial_{\mu}+\Gamma_{\mu}\left(x\right)$, where $\partial_{\mu}$ corresponds to the partial derivative and $\Gamma_{\mu}\left(x\right)=\frac{i}{4}\,\omega_{\mu ab}\left(x\right)\,\Sigma^{ab}$ corresponds to the spinorial connection \cite{naka,bd} and $\Sigma^{ab}=\frac{i}{2}\,\left[\gamma^{a},\,\gamma^{b}\right]$. The matrices $\gamma^{a}$ are the standard Dirac matrices defined in the Minkowski spacetime \cite{greiner}:  
\begin{eqnarray}
\gamma^{0}=\hat{\beta}=\left(
\begin{array}{cc}
1 & 0 \\
0 & -1 \\
\end{array}\right);\,\,\,\,\,\,
\gamma^{i}=\hat{\beta}\,\hat{\alpha}^{i}=\left(
\begin{array}{cc}
 0 & \sigma^{i} \\
-\sigma^{i} & 0 \\
\end{array}\right);\,\,\,\,\,\,\Sigma^{i}=\left(
\begin{array}{cc}
\sigma^{i} & 0 \\
0 & \sigma^{i} \\	
\end{array}\right),
\label{1.8}
\end{eqnarray}
with $\vec{\Sigma}$ being the spin vector. The matrices $\sigma^{i}$ are the Pauli matrices and satisfy the relation $\frac{1}{2}\left(\sigma^{i}\,\sigma^{j}+\sigma^{j}\,\sigma^{i}\right)=\eta^{ij}$. The $\gamma^{\mu}$ matrices are related to the $\gamma^{a}$ matrices via $\gamma^{\mu}=e^{\mu}_{\,\,\,a}\left(x\right)\gamma^{a}$ \cite{bd}. Besides, the term $\omega_{\mu ab}\left(x\right)$ is the spin connection and can be obtained by solving Cartan's structure equations in the absence of torsion $d\hat{\theta}^{a}+\omega^{a}_{\,\,\,b}\wedge\hat{\theta}^{b}=0$ \cite{naka}, where $\omega^{a}_{\,\,\,b}=\omega_{\mu\,\,\,\,b}^{\,\,\,a}\left(x\right)\,dx^{\mu}$ is the 1-form connection, the operator $d$ is the exterior derivative and the symbol $\wedge$ means the wedge product. Hence, by working with the units $\hbar=c=1$ from now on, the Dirac equation in a curved spacetime is written in the form:
\begin{eqnarray}
i\gamma^{\mu}\,\partial_{\mu}\psi+i\gamma^{\mu}\,\Gamma_{\mu}\left(x\right)\psi=m\psi.
\label{1.9}
\end{eqnarray}

In what follows, let us consider a topological defect spacetime background in order to study quantum effects from a Lorentz symmetry breaking effects in the geometrical picture given by the effective metric tensor (\ref{1.7}). The line element of the cosmic string spacetime is written as
\begin{eqnarray}
ds^{2}=dt^{2}-d\rho^{2}-\eta^{2}\rho^{2}d\varphi^{2}-dz^{2}.
\label{1.10}
\end{eqnarray}

The appearance of topological defects in the spacetime, like the cosmic string (\ref{1.10}), is considered to be through phase transitions during the evolution of the Universe which involves a symmetry breaking \cite{td2}. On the other hand, topological defects can be viewed in the solid state context as the analogue of three-dimensional gravity through the formalism proposed by Katanaev and Volovich \cite{kat}. In general, a defect corresponds to singular curvature, torsion or both along the line defect. For instance, the spatial part of the line element (\ref{1.10}) is known in the solid state context as a disclination. The parameter $\eta$ is related to the deficit angle and it is defined as $\eta=1-4\varpi G/c^{2}$, with $\varpi$ being the linear mass density of the cosmic string. The azimuthal angle varies in the interval $0\leq\varphi<2\pi$. The deficit angle can assume only values in which $0<\eta<1$, because values greater than 1 correspond to an anti-cone with negative curvature, which makes sense only in the description of linear defects in solid crystal \cite{kat,furt}. It must be emphasized that this geometry has a curvature tensor which represents a conical singularity $R_{\rho,\varphi}^{\rho,\varphi}=\frac{1-\eta}{4\eta}\,\delta_{2}(\vec{r})$, where $\delta_{2}(\vec{r})$ is the two-dimensional delta function. This behavior of the curvature tensor is denominated conical singularity \cite{staro} because it gives rise to the curvature concentrated on the cosmic string axis, with all other places having null curvature.

Now, let us choose the local reference frame for the observers in the cosmic string spacetime as
\begin{eqnarray}
\hat{\theta}^{a}=dt;\,\,\,\hat{\theta}^{1}=d\rho;\,\,\,\hat{\theta}^{2}=\eta\rho\,d\varphi;\,\,\,\hat{\theta}^{3}=dz,
\label{1.11}
\end{eqnarray}
and let let us consider the normalized parameter four-vector $\xi_{a}$ being a space-like four-vector given by
\begin{eqnarray}
\xi_{a}=\left(0,0,1,0\right).
\label{2.2}
\end{eqnarray}

In this way, we can write the four-vector $\xi_{\mu}\left(x\right)$ given in the effective metric tensor (\ref{1.7}) via relation: $\xi_{\mu}\left(x\right)=e^{a}_{\,\,\,\mu}\left(x\right)\,\xi_{a}=\eta\rho\,\xi_{2}$; thus, we have that condition $\xi_{\mu}\left(x\right)\,\xi^{\mu}\left(x\right)=\mathrm{const}$ established in Refs. \cite{curv,curv2,curv3} is satisfied, and the effective metric of the cosmic string spacetime under Lorentz symmetry breaking effects becomes
\begin{eqnarray}
\bar{ds}^{2}=dt^{2}-d\rho^{2}-\eta^{2}\rho^{2}\left(1+\epsilon\right)d\varphi^{2}-dz^{2}.
\label{2.5}
\end{eqnarray}

From now on, let us establish a new set of vectors which corresponds to the local reference frame of the observers related to the effective metric given in Eq. (\ref{2.5}):
\begin{eqnarray}
\hat{\Theta}^{0}=dt;\,\,\,\hat{\Theta}^{1}=d\rho;\,\,\,\hat{\Theta}^{2}=\eta\rho\sqrt{1+\epsilon}\,\,\,d\varphi;\,\,\,\hat{\Theta}^{3}=dz.
\label{2.6}
\end{eqnarray}

By solving the Cartan structure equations in the absence of torsion $d\hat{\Theta}^{a}+\omega^{a}_{\,\,\,b}\wedge\hat{\Theta}^{b}=0$ \cite{naka}, where $\omega^{a}_{\,\,\,b}=\omega_{\mu\,\,\,\,b}^{\,\,\,a}\left(x\right)\,dx^{\mu}$, we obtain the following non-null components of the 1-form connection $\omega_{\varphi\,\,\,1}^{\,\,\,2}\left(x\right)=-\omega_{\varphi\,\,\,2}^{\,\,\,1}\left(x\right)=\eta\sqrt{1+\epsilon}$. Hence, the only non-null component of the spinorial connection is
\begin{eqnarray}
\Gamma_{\varphi}\left(x\right)=-\frac{i}{2}\,\eta\sqrt{1+\epsilon}\,\Sigma^{3}.
\label{2.7}
\end{eqnarray}

Thereby, the Dirac equation in the cosmic string spacetime under Lorentz symmetry breaking effects becomes
\begin{eqnarray}
m\psi=i\gamma^{0}\,\frac{\partial\psi}{\partial t}+i\gamma^{1}\left(\frac{\partial}{\partial\rho}+\frac{1}{2\rho}\right)\psi+\frac{i\gamma^{2}}{\eta\rho\sqrt{1+\epsilon}}\,\frac{\partial\psi}{\partial\varphi}+i\gamma^{3}\,\frac{\partial\psi}{\partial z}.
\label{2.8}
\end{eqnarray}

By applying the Dirac phase factor method \cite{dirac,dirac2} into the Dirac equation (\ref{2.8}), where we can write the wave function in the form:
\begin{eqnarray}
\psi=e^{i\phi}\,\psi_{0},
\label{2.9}
\end{eqnarray}
where $\psi_{0}$ is the solution of the following equation
\begin{eqnarray}
m\psi_{0}=i\gamma^{0}\,\frac{\partial\psi_{0}}{\partial t}+i\gamma^{1}\,\frac{\partial\psi_{0}}{\partial\rho}+\frac{i\gamma^{2}}{\eta\rho\sqrt{1+\epsilon}}\,\frac{\partial\psi_{0}}{\partial\varphi}+i\gamma^{3}\,\frac{\partial\psi_{0}}{\partial z},
\label{2.10}
\end{eqnarray}
and the relativistic geometric phase acquired by the wave function of the Dirac particle is
\begin{eqnarray}
\phi=\frac{1}{2}\oint\eta\sqrt{1+\epsilon}\,\Sigma^{3}\,d\varphi=\pi\,\eta\,\sqrt{1+\epsilon}\,\,\Sigma^{3}.
\label{2.11}
\end{eqnarray}

Hence, without applying the adiabatic approximation we have obtained the geometric quantum phase acquired by the wave function of a Dirac particle which consists in a Aharonov-Anandan quantum phase \cite{ahan}. This relativistic geometric quantum phase is yielded by the effects of the Lorentz symmetry breaking defined by the normalized parameter four-vector given in Eq. (\ref{2.2}) and the topology of the cosmic string spacetime. Two characteristics of the relativistic geometric phase (\ref{2.11}) is that it is a non-Abelian geometric phase \cite{anan,anan2,anan3} and it does not depend on the velocity of the particle, that is, it is non-dispersive phase \cite{disp,disp2,disp3}. 

Moreover, by taking $\epsilon=0$, then, we recover the results obtained in Refs. \cite{bf2,bf3} which correspond to the relativistic geometric quantum phase yielded only by the topology of the cosmic string spacetime. On the other hand, by taking the limit $\eta\rightarrow1$, we obtain the Minkowski spacetime under the influence of Lorentz symmetry breaking effects defined by the normalized parameter four-vector given in Eq. (\ref{2.2}). In this case, the relativistic geometric phase (\ref{2.11}) is yielded only by the effects of the violation of the Lorentz symmetry.

\section{nonrelativistic geometric quantum phase}

An interesting question is on the meaning of the nonrelativistic behaviour of a fermion in a fully relativistic background. It is well-known that the spin of a particle is a property that appears naturally in the relativistic description. Thereby, an accurate way of investigating the kinematic properties of a particle under the influence of a background that causes the anisotropy of the spacetime is via the Dirac equation. However, if we intend evaluate these contributions that arise from high energy in the laboratory, we need to investigate these effects in the nonrelativistic limit. At present days, there is a line of research in which bounds related to the intensity of this background are evaluated by analysing the nonrelativistic limit of the Dirac equation \cite{baeta,belich,belich1,belich3}. In this section, our goal is to obtain the nonrelativistic geometric phase for a spin-$1/2$ particle yielded by the Lorentz symmetry breaking effects defined by the normalized parameter four-vector given in Eq. (\ref{2.2}) in the presence of a topological defect. Let us discuss the nonrelativistic behaviour of the spin-$1/2$ neutral particle by using the Foldy-Wouthuysen approximation \cite{fw,greiner}. In this approach, we need first to write the Dirac equation in the form:
\begin{eqnarray}
i\,\frac{\partial\psi}{\partial t}=\hat{H}\,\psi,
\label{3.1}
\end{eqnarray} 
where the Hamiltonian operator $\hat{H}$ of the system must be written in terms of even operators $\hat{\mathcal{E}}$ and odd operators $\hat{\mathcal{O}}$ as follows:
\begin{eqnarray}
\hat{H}=m\,\hat{\beta}+\hat{\mathcal{E}}+\hat{\mathcal{O}}.
\label{3.2}
\end{eqnarray}
The even and odd operators written in (\ref{3.2}) must satisfy the following relations: 
\begin{eqnarray}
\left[\hat{\mathcal{E}},\hat{\beta}\right]&=&\hat{\mathcal{E}}\,\hat{\beta}-\hat{\beta}\,\hat{\mathcal{E}}=0;\nonumber\\
[-2mm]\label{3.3}\\[-2mm]
\left\{\hat{\mathcal{O}},\hat{\beta}\right\}&=&\hat{\mathcal{O}}\,\hat{\beta}+\hat{\beta}\,\hat{\mathcal{O}}=0.\nonumber
\end{eqnarray} 

The aim of the Foldy-Wouthuysen approach \cite{fw,greiner} is to apply a unitary transformation in order to remove the operators from the Dirac equation that couple the ``large'' to the ``small'' components of the Dirac spinors. Basically, the even operators $\hat{\mathcal{E}}$ do not couple the ``large'' to the ``small'' components of the Dirac spinors, while the odd operators $\hat{\mathcal{O}}$ do couple them. Moreover, we can see that both even and odd operators must be Hermitian operators. Thereby, by applying the Foldy-Wouthuysen approximation \cite{fw,greiner} up to the terms of order $m^{-1}$, we can write the nonrelativistic limit of the Dirac equation in the form:
\begin{eqnarray}
i\frac{\partial\psi}{\partial t}=m\hat{\beta}\psi+\hat{\mathcal{E}}\psi+\frac{\hat{\beta}}{2m}\,\hat{\mathcal{O}}^{2}\psi.
\label{3.4}
\end{eqnarray}

Our first goal is to write the Dirac equation (\ref{2.8}) in the form given in (\ref{3.1}) and, then, identify all even and odd operators as established previously. Hence, after some calculations, we have 
\begin{eqnarray}
\hat{\mathcal{O}}=\vec{\alpha}\cdot\vec{\pi};\,\,\,\,\hat{\mathcal{E}}=0,
\label{3.5}
\end{eqnarray}
where $\vec{\pi}=\vec{p}-i\vec{\xi}$ and the vector $\vec{\xi}$ has the components given by $-i\xi_{k}=-\frac{\sigma^{3}}{2\rho}\,\delta_{2k}$. Note that the vector $\vec{\xi}$ arises from the spinorial connection $\Gamma_{\mu}\left(x\right)$ discussed above \cite{bf3,bbs3,bb,lbb}. Thereby, substituting (\ref{3.5}) into (\ref{3.4}), we obtain the nonrelativistic limit of the Dirac equation (\ref{2.8}). Henceforth, we work with two-spinors, then, the corresponding Schr\"odinger-Pauli equation is
\begin{eqnarray}
i\frac{\partial\psi}{\partial t}=m\,\hat{\beta}\,\psi+\frac{\hat{\beta}}{2m}\left[\vec{p}-i\vec{\xi}\right]^{2}\psi.
\label{3.6}
\end{eqnarray}

By applying the Dirac phase factor method \cite{dirac,dirac2} into the Schr\"odinger-Pauli equation (\ref{3.6}), where we can decompose the wave function in the same way as done in Eq. (\ref{2.9}), we have in this case that $\psi_{0}$ is the solution of the following equation (for two-spinors):
\begin{eqnarray}
i\frac{\partial\psi_{0}}{\partial t}=-\frac{1}{2m}\,\nabla^{2}\psi_{0},
\label{3.7}
\end{eqnarray}
and the geometric quantum phase acquired by the wave function of the nonrelativistic spin-$1/2$ particle is
\begin{eqnarray}
\phi=\frac{1}{2}\oint\eta\sqrt{1+\epsilon}\,\sigma^{3}\,d\varphi=-\pi\,\eta\,\sqrt{1+\epsilon}\,\,\sigma^{3}.
\label{3.8}
\end{eqnarray}

Hence, we have obtained a geometric quantum phase in the nonrelativistic case without applying the adiabatic approximation, therefore the phase shift given in Eq. (\ref{3.8}) is a Aharonov-Anandan quantum phase \cite{ahan}. As in the relativistic case, the nonrelativistic geometric phase (\ref{3.8}) possesses two characteristics: it is a non-Abelian phase \cite{anan,anan2,anan3} and a non-dispersive phase \cite{disp,disp2,disp3}. The phase shift (\ref{3.8}) is yielded by the effects of the Lorentz symmetry violation defined by the normalized parameter four-vector given in Eq. (\ref{2.2}) and the topology of a defect. By comparing with the solid state context, this second contribution to the nonrelativistic geometric phase arises from the presence of linear topological defect called disclination \cite{kat}.

Besides, by taking $\epsilon=0$, we have that the geometric quantum phase (\ref{3.8}) is yielded by the topology of the defect, which recover the results of Refs. \cite{bf2,bf3}. However, by taking the limit $\eta\rightarrow1$, the geometric phase (\ref{3.8}) is yielded by the Lorentz symmetry breaking effects defined by the normalized parameter four-vector given in Eq. (\ref{2.2}) without the influence of the defect.

\section{conclusion}

We have analysed the arising of geometric quantum phases in the relativistic and nonrelativistic quantum dynamics of a Dirac neutral particle from the effects of the violation of the Lorentz symmetry in the cosmic string spacetime. By starting from the modified Maxwell theory coupled to gravity \cite{curv2,curv3}, we have established a spacelike normalized parameter four-vector (see Eq. (\ref{2.9})) and written an effective metric for the cosmic string spacetime under Lorentz symmetry breaking effects. Then, without applying the adiabatic approximation, we have shown that the wave function of a Dirac particle acquires a non-Abelian and non-dispersive relativistic geometric quantum phase yielded by the effects of the Lorentz symmetry breaking defined by the normalized parameter four-vector given in Eq. (\ref{2.2}) and the topology of the cosmic string spacetime.

Moreover, we have shown by taking the limit $\eta\rightarrow1$ that the relativistic geometric phase is yielded by the effects of the violation of the Lorentz symmetry in the Minkowski spacetime. On the other hand, by taking $\epsilon=0$, the Lorentz symmetry breaking effects vanish and we recover the relativistic geometric quantum phase yielded only by the topology of the cosmic string spacetime obtained in Refs. \cite{bf2,bf3}.

We have also discussed the nonrelativistic limit of the Dirac equation by applying the Foldy-Wouthuysen approximation \cite{fw,greiner}. We have show that the wave function of a nonrelativistic spin-$1/2$ particle acquires a geometric quantum phase without using the adiabatic approximation \cite{ahan}. This phase shift corresponds to a non-Abelian and non-dispersive geometric phase which is yielded by the effects of the Lorentz symmetry violation defined by the normalized parameter four-vector given in Eq. (\ref{2.2}) and the topology of a defect. Furthermore, we have shown by taking the limit $\eta\rightarrow1$ that the nonrelativistic geometric phase is yielded only by the Lorentz symmetry breaking effects defined by the normalized parameter four-vector given in Eq. (\ref{2.2}) without the influence of the defect. However, by taking $\epsilon=0$, the nonrelativistic geometric phase is yielded only by the topology of the defect.

It is interesting to note that the arising of geometric quantum phases for fermions coupled to a curved spacetime background can be useful in studies of particles in the vicinity of black holes and in scenarios of the Lorentz symmetry violation. Another theme in which has not been investigated yet is the Lorentz symmetry violation in a supergravity scenario. Besides, if we consider charged fermions, we can introduce the electromagnetic minimal coupling, then, new results can be expected such as new contributions to the Aharonov-Bohm effect can arise from the scenario of the violation of the Lorentz symmetry. We hope to bring this discussion in the near future.

\acknowledgments{The authors would like to thank CNPq (Conselho Nacional de Desenvolvimento Cient\'ifico e Tecnol\'ogico - Brazil) for financial support.}

\end{document}